\documentclass[conference]{IEEEtran}
\usepackage{cite}
\usepackage{amsmath,amssymb,amsfonts}
\usepackage{algorithmic}
\usepackage{graphicx}
\usepackage{textcomp}
\usepackage{xcolor}
\usepackage{comment}
\usepackage{authblk}
\usepackage[hyphens]{url}

\def\BibTeX{{\rm B\kern-.05em{\sc i\kern-.025em b}\kern-.08em
    T\kern-.1667em\lower.7ex\hbox{E}\kern-.125emX}}

\title{RETROSPECTIVE: Corona: System Implications of Emerging Nanophotonic Technology}

\author[ ]{Dana Vantrease}
\author[1]{Robert Schreiber}
\author[2]{Matteo Monchiero}
\author[3]{Moray McLaren}
\author[3]{\\Norman P. Jouppi}
\author[4]{Marco Fiorentino}
\author[5]{Al Davis}
\author[4]{Raymond G. Beausoleil}
\author[7]{Jung Ho Ahn}

\affil [ ] { 
\textsuperscript{1} Cerebras \quad  \textsuperscript{2} Snowflake 
\quad \textsuperscript {3} Google 
\quad \textsuperscript{4} Hewlett Packard Enterprise  
\quad \textsuperscript{5} Nvidia 
\quad \textsuperscript{7} Seoul National University 
}

\affil[ ]{\textit{rob.schreiber@cerebras.net, matteo.monchiero@snowflake.com, moray@google.com, }}
\affil [ ] {\textit{jouppi@google.com, marco.fiorentino@hpe.com, ald@nvidia.com, ray.beausoleil@hpe.com, gajh@snu.ac.kr}}

\begin{document}
\maketitle
\thispagestyle{plain}
\pagestyle{plain}




\section{An Overview: What We Were Trying to Do} 


The 2008 Corona effort~\cite{2008-isca-corona} was inspired by a pressing need for more of everything, as demanded by the salient problems of the day.  Dennard scaling was no longer in effect.  A lot of computer architecture research was in the doldrums.  Papers often showed incremental subsystem performance improvements, but at incommensurate cost and complexity.  The many-core era was moving rapidly, and the approach with many simpler cores was at odds with the better and more complex subsystem publications of the day.  Core counts were doubling every 18 months, while per-pin bandwidth was expected to double, at best, over the next decade.  Memory bandwidth and capacity had to increase to keep pace with ever more powerful multi-core processors.  With increasing core counts per die, inter-core communication bandwidth and latency became more important.  At the same time, the area and power of electrical networks-on-chip were increasingly problematic: To be reliably received, any signal that traverses a wire spanning a full reticle-sized die would need significant equalization, re-timing, and multiple clock cycles.  This additional time, area, and power was the crux of the concern, and things looked to get worse in the future.  

New packaging and integration options were emerging in response to these problems.  The end of Dennard scaling pushed 2.5D and 3D packaging technologies.  This allowed diverse process technologies such as DRAM, digital logic, and analog I/O drivers to be developed in the best process and then co-packaged.  This trend is evident even today, where logic and main-memory processes remain starkly different.  More importantly, TSMC, Intel, and Samsung have all announced that their 2nm processes, targeted for production in 2025, will only provide short chip-to-chip I/O capability.  Driving circuit board traces will be left to other devices, presumably co-packaged.

Silicon nanophotonics was of particular interest and seemed to be improving rapidly.   This led us to consider taking advantage of 3D packaging, where one die in the 3D stack would be a photonic network layer.  The electrical communications would be sub-mm scale within a processing tile, with short, vertical through-silicon vias to an analog driver layer providing the interface between the logic tiles and the photonic network layer for inter-tile and off die communication.  We chose a waveguide-per-destination approach, avoiding any need for on-die switches that would only add to the power, area, and delay headache.   

Our focus was on a system that could be built about a decade out.  Thus, we tried to predict how the technologies and the system performance requirements would converge in about 2018.  Corona was the result this exercise; now, 15 years later, it's interesting to look back at the effort.


\section{Origins} 

In 2007, HP Labs began investigating whether the silicon photonic interconnect technology then under development in research labs could meet the bandwidth demands of future many-core systems.  Photonics had already proven itself as a low-power, low-latency communication medium over meter, or greater, distances.  We asked whether on-chip or on-stack nanophotonics would be a practical and beneficial alternative to electronics for cross-chip communication as chip sizes continued to grow. 


A multi-disciplinary team of ten HP architects and physicists collaborated closely on the Corona design, a 3D stacked 256-core architecture with nanophotonic on- and off-chip interconnects.  It started with a blank slate.  The architects worked backward, deriving bandwidth and power requirements from future compute projections.  Meanwhile, the photonics engineers 
projected the photonic technology forward.  Meeting in the middle, we aimed for a realistic and holistically sound system design.  For example, applying photonics on-chip just pushed the bandwidth problem off-chip, so we applied photonics off-chip, too.   We also built an optical arbiter that could keep pace with the optical interconnect.  The Corona architecture was novel.  Multiple patents were awarded for this work.  We opine that the design would have been practical had our technology maturity predictions been 
correct. 


The silicon-photonic technology of 2008 was promising but also immature.  Without regrets, we invested in designing a system that was never fabricated.  The team stretched to learn new fields and invented diction to communicate, e.g., ``modulating rainbows'' and ``snaking wave guides.''  Each day we gathered with colored dry-erase markers and Matlab models in view of Bill Hewlett and David Packard's Palo Alto offices -- the ones they used after moving out of the garage.  In the company of top-notch colleagues and with the HP innovation legacy, it was easy to dream big.

\section{Impact} 

To understand the impact of a radical new way to communicate across the chip,
we had to look at a comprehensive system architecture rather than focus on improving individual subsystems.  We also had to consider basic infrastructure issues like power, cooling, and packaging.  While we relied heavily on emerging technologies, we needed to be reasonably confident that these technologies could mature a decade or so in the future.  We based our design and analysis on demonstrable laboratory hardware prototypes and technology improvement rate projections (e.g. CMOS, photonics, and memory).

So how did we do?  We got a lot of things right: core counts would continue to explode, bandwidth requirements and memory capacity would grow commensurately, and optical interconnect would be an important component of future high-performance systems.  There were places where we were overly optimistic.  Other things we just missed, and now two areas seem to be the most important.  First, the rate at which data centers would grow into the central role they now play has proved to be a more emphatic forcing function for architecture than core count growth.  
Second, we overshot on photonic technology's maturity with regard to physical integration, temperature sensitivity, and communication density.

Data centers became foundational about when we wrote the paper and smart phones were becoming ubiquitous.  Modern mobile phones have stunning capabilities but without the data center, many of their capabilities would be missing.  Today the largest data center is over 10 million square feet. In these massive data centers, the move to optical interconnect is mandatory.  Moreover, AI now dominates the computational landscape --- where 
high performance computing (HPC) was at the leading edge of computing, AI has now displaced it.   A different breed of highly parallel chips is emerging to meet the more regular and fine-grain parallel needs of AI, chips and systems like the Google TPU, the Cerebras wafer-scale engine, the computational pipeline of the SambaNova, and others.   None is a shared-memory multi-core design.

We were too optimistic about how rapidly silicon photonics would mature, and perhaps too pessimistic about electrical alternatives, when we 
proposed that photonic interconnects would be pushed down to a very fine-grain (sub-mm) level.     For Corona, we employed a reticle-sized 
silicon photonic interposer that connected 64 quad-processor clusters.  Reticle-sized photonic interposers do not exist today.  
A major reason for this
is the difficulty in doing fiber attach to the photonic substrate.  
However, larger-than-reticle sized organic and silicon interposers are common in high-end data 
center packaging schemes, which mostly use easier to cool 2.5D packaging rather than 3D approaches.
3D packaging with fine-grained photonics, like that of Corona, has not happened. The temperature sensitivity, reliability, and heat density of modern computing dies continue to be a barrier.


Finally, we are also a long way from deploying 64 wavelengths of light in a single fiber in the dense wave division multiplexing scheme that we employed in Corona.  In the past 15 years, most silicon photonics companies have used a \emph{vertical} business model where they have tried to control all aspects of their technology, from design tools to packaging.  This has prevented the development of a robust solution ecosystem and the kind of investments needed to enable the 3D packaging technologies we envisioned for Corona.  In short, we bundled the right technologies but were too optimistic about how quickly quality photonic components would be available.  

We did not know the fate of photonic interconnects when we designed Corona, and so it is encouraging that photonics is standard in today's data centers.  Inter-rack and longer optical communication with pluggable Active Optical Cables (AOCs) are commonplace.  Optical Circuit Switches (OCS) use mirrors for all-optical, microsecond-reconfigurable switching.  
While cost and reliability continue to be pressing issues, the technology is progressing and its future \emph{still} looks promising.

\section{Conclusions} 

We learned a lot doing this work.  Now we have the luxury of 20-20 hindsight to 
see how well our ability to predict the future worked out.  No surprise, our crystal ball wasn't perfect, as we have noted above.  But the effort was exciting, rewarding, and elucidating.  As is often the case, we learned from our mistakes.

Allow us to wrap up with some advice for future researchers, and particularly for grad students: Beware of low-hanging fruit.  It may result in a publication and academic brownie points.  However, the long-term significance of incremental and obvious solutions is likely to be small.   Don't waste time on the easy stuff.  Take risks, aim high, and choose problems that will make a difference if you solve them. People will find your work interesting, even compelling, and motivating.  What architect Daniel Burnham said in 1910 about his booming city of Chicago also applies to computer architects: ``Make no little plans, they have no magic in them to stir [the] blood.''  We hope that the Corona work will continue to stir the blood of future computer engineering researchers.  

\section*{Acknowledgements}
While writing this retrospective, we have had in our thoughts our co-author 
Nate Binkert, who died in 2017.  We could not have asked for a better person to work with.  He will always be missed.  

We would also like to thank HP Laboratories for giving us the freedom to focus on a ``concept'' architecture of the future rather than limiting our ambitions to what was safe.


\bibliographystyle{IEEEtran}
\bibliography{ref}

\end{document}